\documentclass[9pt]{sig-alternate-arxiv}
\usepackage[latin1]{inputenc}
\usepackage{longtable,helvet,courier,fancyhdr}
\usepackage{amsfonts}
\usepackage{alltt}
\usepackage{chemarr}
\usepackage[version=3]{mhchem}
\usepackage{amsmath,amssymb,amsfonts,bbm}
\usepackage[linesnumbered,tworuled,vlined]{algorithm2e}
\usepackage{enumerate}
\newif\ifarxiv

\usepackage[dvipsnames]{xcolor}

\usepackage{graphicx}

\newcommand{\VS}{\operatorname{VS}}

\def \R{\mathbb{R}}
\def \Z{\mathbb{Z}}

\newcommand{\RRp}{\left]0,\infty\right[}

\newcommand{\difft}[1]{\dot{#1}}

\newcommand{\phmax}{\phantom{\max\mathopen(\kern2pt}}

\newcommand{\hquad}{\kern0.5em}

\newcommand{\bemerkung}[1]{\relax}

\newlength{\lrtskip}
\begin{document}
\setcopyright{acmcopyright}

\makeatletter
\@printcopyrightfalse
\makeatother

\title{A Case Study on the Parametric Occurrence of\\ Multiple Steady States}

\numberofauthors{2}

\author{
%
%
%
  \alignauthor Russell Bradford\\
  \affaddr{
  University of Bath, U.K.}\\
  \email{R.J.Bradford@bath.ac.uk}  
  \and
  \alignauthor James H.Davenport\\
  \affaddr{
    University of Bath, U.K.}\\
  \email{J.H.Davenport@bath.ac.uk}  
  \and
  \alignauthor Matthew England\\
  \affaddr{
  Coventry University, U.K.}\\
  \email{Matthew.England@coventry.ac.uk}
  \and
  \alignauthor Hassan Errami\\
  \affaddr{
  University of Bonn, Germany}\\
  \email{errami@cs.uni-bonn.de}
  \and
  \alignauthor Vladimir Gerdt\\
  \affaddr{JINR, Dubna, Russia}\\
  \email{gerdt@jinr.ru}
  \and
  \alignauthor Dima Grigoriev\\
  \affaddr{CNRS \& University of Lille, France}\\
  \email{dmitry.grigoryev@math.univ-lille1.fr}
  \and
  \alignauthor Charles Hoyt\\
  \affaddr{b-it, Bonn, Germany}\\
  \email{cthoyt@gmail.com}
  \and
  \alignauthor Marek Ko{\v s}ta\\
  \affaddr{
    Slovak Academy of Sciences}\\
  \email{marek.kosta@savba.sk}
  \and
  \alignauthor Ovidiu Radulescu\\
  \affaddr{DIMNP UMR CNRS/UM 5235,\\
  University of Montpellier, France}\\
  \email{ovidiu.radulescu@umontpellier.fr}
  \and
  \alignauthor Thomas Sturm\\
  \affaddr{U Lorraine, CNRS, Inria \& LORIA, Nancy, France}\\
  \affaddr{MPI Informatics \& Saarland University, Germany}\\
  \email{thomas.sturm@loria.fr}
  \and
  \alignauthor
  Andreas Weber\\
  \affaddr{University of Bonn, Germany}\\
  \email{weber@cs.uni-bonn.de}
  \and
  \alignauthor
}


\maketitle

\begin{abstract}

We consider the problem of determining multiple steady states for positive real values in models of biological networks.  Investigating the potential for these in models of the mitogen-acti\-vated protein kinases (MAPK) network has consumed considerable effort using special insights into the structure of corresponding models.
Here we apply combinations of symbolic computation methods for mixed equality/inequality systems, specifically virtual substitution, lazy real triangularization and cylindrical algebraic decomposition.  We determine multistationarity of an 11-dimensional MAPK network when numeric values are known for all but potentially one parameter.  More precisely, our considered model has 11 equations in 11 variables and 19 parameters, 3 of which are of interest for symbolic treatment, and furthermore positivity conditions on all variables and parameters.

\end{abstract}

\sloppy

\section{Introduction}

The occurrence of multiple steady states (multistationarity) has important
consequences on the capacity of signaling pathways to process biological
signals, even in its elementary form of two stable steady states (bistability).
Bistable switches can act as memory circuits storing the information needed for
later stages of processing \cite{WengBhallaLyengar1999}. The response of
bistable signaling pathways shows hysteresis, namely dynamic and static lags between input and output. Because of hysteresis one can have, at the same time, a sharp binary response and protection against chatter noise.

Bistability of signaling usually occurs as a result of activation of upstream signaling proteins by downstream components \cite{BhallaLyengar99}.
A different mechanism for producing bistability in signaling pathways was proposed by Markevich et al. \cite{Markevich2004}.
In this mechanism bistability can be caused by multiple phosphorylation/dephosphorylation cycles that share enzymes.
A simple, two-step phosphorylation/dephosphorylation cycle is capable of ultrasensitivity, a form of all or nothing
response with no hysteresis (Goldbeter--Koshland mechanism).
In multiple phosphorylation/dephosphorylation cycles, enzyme sharing provides competitive interactions and positive feedback that ultimately leads to bistability.

Algorithmically the task is to find the positive real solutions of a parameterized
system of polynomial or rational systems, since  the dynamics of the network is
given by polynomial systems (arising from mass action kinetics) or rational functions (arising in signaling networks when some intermediates of the reaction mechanisms are reduced).
Due to the high computational complexity of this task \cite{GrigorevVorobjov1988a}
 considerable work has been done to use specific properties of networks and to
 investigate the potential of multistationarity of a biological network
 out of the network structure.  
This only determines whether or not there exist rate constants allowing multiple steady states, instead of coming up with a semi-algebraic description of the range of parameters yielding this property.
These approaches can be traced back to the origins of Feinberg's {\em Chemical Reaction Network Theory} (CRNT)
whose main result is that networks of deficiency 0 have a unique positive steady state for all rate constants
\cite{Feinberg1987,CraciunDickensteinShiuSturmfels2009}.
We refer to \cite{Conradi2008,PerezMillan2015,Johnston2014} for the use of CRNT and other graph theoretic methods to determine potential existence of multiple positive steady states, with \cite{JoshiShiu2015} giving a survey.

Given a bistable mechanism it is also important to compute the bistability domains in parameter space,
namely the parameter values for which there is more than one stable steady state.
The size of bistability domains gives the spread of the hysteresis and quantifies the robustness of the switches.
The work of  Wang and Xia \cite{WangXia2005a} is relevant here: they used symbolic computation tools, including cylindrical algebraic decomposition as we do below, to determine the
number of steady states and their stability for several systems.  They reported results up to a 5-dimensional system using specified
parameter values, but their method is extensible to parametric questions. 
Higher-dimensional systems were studied using sign conditions on the coefficients of the characteristic polynomial of the Jacobian. In some cases these guarantee uniqueness of the steady state \cite{ConradiMincheva}.

In this paper we use an 11-dimensional model of a  mitogen-activated protein kinases
(MAPK) cascade \cite{Markevich2004} as a case study to investigate properties of the system using algorithmic methods towards the goal of semi-algebraic descriptions of parameter
regions for which multiple positive steady states exist.

\section{The MAPK Network} 
\label{secMAPK:system}

The model of the MAPK cascade we are investigating can be found in the Biomodels
database \cite{Li2010a}.\footnote{\url{www.ebi.ac.uk/biomodels-main/BIOMD0000000026}}
We have renamed the species names to $x_1$, \dots,~$x_{11}$ and the rate
constants to $k_1$, \dots,~$k_{16}$ to facilitate reading:
\begin{align}
  %
  \difft{x_1} ={} &  k_{2} x_{6} + k_{15} x_{11} - k_{1} x_{1} x_{4} - k_{16} x_{1} x_{5}\nonumber\\
  \difft{x_2} ={} &  k_{3} x_{6} + k_{5} x_{7} + k_{10} x_{9} + k_{13} x_{10} - \nonumber\\
     &  \quad x_{2} x_{5} (k_{11} + k_{12}) - k_{4} x_{2} x_{4}\nonumber\\
  \difft{x_3} ={} &  k_{6} x_{7} + k_{8} x_{8} - k_{7} x_{3} x_{5}\nonumber\\
  \difft{x_4} ={} &  x_{6} (k_{2} + k_{3}) + x_{7} (k_{5} + k_{6}) - k_{1} x_{1} x_{4} - k_{4} x_{2} x_{4}\nonumber\\
  \difft{x_5} ={} &  k_{8} x_{8} + k_{10} x_{9} + k_{13} x_{10} + k_{15} x_{11} - \nonumber\\
    &   \quad  x_{2} x_{5} (k_{11} + k_{12}) - k_{7} x_{3} x_{5} - k_{16} x_{1} x_{5}\nonumber\\
  \difft{x_6} ={} &  k_{1} x_{1} x_{4} - x_{6} (k_{2} + k_{3})\nonumber\\
  \difft{x_7} ={} &  k_{4} x_{2} x_{4} - x_{7} (k_{5} + k_{6})\nonumber\\
  \difft{x_8} ={} &  k_{7} x_{3} x_{5} - x_{8} (k_{8} + k_{9})\nonumber\\
  \difft{x_{9}} ={} &  k_{9} x_{8} - k_{10} x_{9} + k_{11} x_{2} x_{5}\nonumber\\
  \difft{x_{10}} ={} &    k_{12} x_{2} x_{5} - x_{10} (k_{13} + k_{14})\nonumber\\
  \difft{x_{11}} ={} &    k_{14} x_{10} - k_{15} x_{11} + k_{16} x_{1} x_{5}.\label{EQ:thesystem}
\end{align}
The Biomodels database also gives us meaningful values for the
rate constants:
\begin{align}
  k_{1} &= 0.02,&\!
  k_{2} &= 1,&\!
  k_{3} &= 0.01,&\!
  k_{4} &= 0.032,\nonumber\\
  k_{5} &= 1,&\!
  k_{6} &= 15,&\!
  k_{7} &= 0.045,&\!
  k_{8} &= 1,\nonumber\\
  k_{9} &= 0.092,&\!
  k_{10} &= 1,&\!
  k_{11} &= 0.01,&\!
  k_{12} &= 0.01,\nonumber\\
  k_{13} &= 1,&\!
  k_{14} &= 0.5,&\!
  k_{15} &= 0.086,&\!
  k_{16} &= 0.0011.\label{EQ:rcestimates}
\end{align}
Some of these values are measured and some are well-educated guesses.  For the purpose of our study we assume they are suitable.

We add three linear conservation constraints introducing three further constant parameters $k_{17}$, $k_{18}$, $k_{19}$:
\begin{align}
  x_{5}  + x_{8} + x_{9} + x_{10} + x_{11} &= k_{17}\nonumber\\
  x_{4} + x_{6} + x_{7} &=  k_{18}\nonumber\\
  x_{1} + x_{2} + x_{3} + x_{6} + x_{7} + x_{8} + x_{9} + x_{10} + x_{11} &=  k_{19}.\label{EQ:claws}
\end{align}
Computations to produce these in MathWorks SimBiology use the left-null space of the stoichiometric matrix under positivity conditions, see for example \cite{schuster1991determining}.

Meaningful values for $k_{17}, k_{18}, k_{19}$ are harder to obtain than the constants in (\ref{EQ:rcestimates}).  The following are some realistic values estimated by ourselves on the basis of our understanding of the biological model:
\begin{align}
  k_{17} &= 100,&  k_{18} &= 50,& k_{19}\in\{200,500\}.\label{EQ:clestimates}
\end{align}
The long-term goal of our research is to treat all three of these together parametrically, although in the present work we focus on situations with one free-parameter.

The steady state problem for the MAPK cascade can now be formulated as a real algebraic problem. 
That is, we replace the left hand sides of all equations in (\ref{EQ:thesystem}) with $0$. This together with the equations in
(\ref{EQ:claws}) yields an algebraic system with polynomials in
\begin{displaymath}
F\subset\Z[k_1,\dots,k_{19}][x_1,\dots,x_{11}].
\end{displaymath}
All entities in our model are strictly positive, which
yields in addition a system
$$P=\{k_{1},\dots,k_{19},x_1,\dots,x_{11}\}\subset\Z[k_1,\dots,k_{19}][x_1,\dots,x_{11}]$$
establishing a side condition on the solutions of $F$ that all variables $x_i$ and parameters $k_i$ of $P$ be positive. In terms of first-order logic our specification of $F$ and $P$
yields a quantifier-free Tarski formula
\begin{equation}
\label{eq:varphi}
\varphi=\bigwedge_{f\in F}f=0\land\bigwedge_{v\in P}v>0.
\end{equation}
The estimations for the rate constants in (\ref{EQ:rcestimates}) formally
establish a substitution rule $\sigma = [0.02/k_1, \dots, 0.0011/k_{16}]$, which can be applied to $F$, $P$, or $\varphi$ in
postfix notation.

\subsection{Symbolic Determination of Occurrences of Multiple Steady States}

In this section we are going to analyze the system for multiple positive steady
states. 
As we will not include a priori information about the stability of the fixed points,
we do not only have to consider (at least) two stable fixed points but also unstable fixed points, i.e.,
we investigate the existence of at least three different roots $\mathbf{x}\in\RRp^{11}$ of
$F$ for given choices $\mathbf{k}\in\RRp^{19}$ of parameters.

We present two investigations: one using the Redlog package in Reduce and the other using the  Regular Chains Library in Maple.  
Both will make use of Cylindrical Algebraic Decomposition (CAD) \cite{ACM:84} to solve the problem.  
The worst-case time complexity of 
CAD is doubly exponential.
\footnote{\label{fn:complexity}Traditionally, doubly exponential in the number of variables.
However recent progress  on CAD in the presence of  equational constraints \cite{EBD15}, such as (\ref{EQ:thesystem}) with $0$ for left-hand side,
 allows us to conclude it is actually doubly-exponential the 
 number of variables minus the number of 
 equational constraints at different levels 
 of the projection \cite{ED16a}.}  
Our approaches admit, in principle, arbitrary numbers of indeterminates. However, for the sake of realistic computation times we must restrict ourselves to one free parameter.  Even then, the number of variables present is too large for contemporary CAD implementations.  We make progress by combining CAD with additional symbolic methods.  Our first approach uses virtual substitution techniques and the second real triangularization.  In both cases we have combined the corresponding methods by hand, but automation is clearly possible.

\subsubsection{Real Quantifier Elimination in Redlog}
Real Quantifier Elimination (QE) can directly handle the parametric existence of steady states, taking as input $\exists x_1\dots\exists x_{11}\varphi$, possibly with substitutions for some parameters. However, we are not only interested in the existence but also in the number
of solutions. We are going to combine Virtual Substitution (VS)
\cite{Weispfenning:97b} with CAD. The former smoothly eliminates the majority of the
quantifiers while the latter allows us to count numbers of solutions via
decomposition of the remaining low-dimensional spaces. That combination of
methods requires the solution of several QE runs with each problem and some
combinatorial arguments. Throughout this subsection we are using
Redlog~\cite{DolzmannSturm:97a}.

\paragraph{Parameter-Free Computations}
We consider $\varphi_{500}=\varphi\sigma[100/k_{17},50/k_{18},500/k_{19}]$ where
all parameters have been substituted with rational numbers. The closed formula
$\bar\varphi_{500}=\exists x_1\dots\exists x_{11}\varphi_{500}$ states the
existence of a suitable real solution. In a first step, we solve for
$i\in\{1,\dots,11\}$ the following eleven QE problems using VS:
\begin{displaymath}
  \varphi_{500}^{(i)}=\VS(\exists x_1\dots\exists x_{i-1}\exists
  x_{i+1}\dots\exists x_{11}\varphi_{500}).
\end{displaymath}
Each $\varphi_{500}^{(i)}$ is a univariate quantifier-free formula describing
all possible real choices for $x_i$ for which there exist real choices for all
other variables such that $\varphi_{500}$ holds. CAD can easily decompose the
corresponding one-dimensional spaces. It turns out that for each $x_i$ there are
exactly three zero-dimensional cells $a_i$, $b_i$, $c_i\in\R$ for which
$\varphi_{500}^{(i)}$ holds. We extract all $a_i$, $b_i$, and $c_i$ as
\emph{real algebraic numbers}, i.e., univariate defining polynomials with
integer coefficients plus
isolating intervals. By combinatorial arguments it is not hard to see that the
following holds for the set $S_{500}$ of real solutions of $\varphi_{500}$:
\begin{displaymath}
  3\leq|S_{500}|
  \quad\text{and}\quad
  S_{500}\subseteq \textstyle \prod_{i=1}^{11}\{a_i,b_i,c_i\}.
\end{displaymath}
Notice that at this point we have proven multistationarity for $k_{19}=500$. We
can furthermore compute $S_{500}$ by plugging the $3^{11}$ candidates from the
Cartesian product into $\varphi_{500}$. A straightforward approach requires
arithmetic with real algebraic numbers followed by the determination of the
signs of the results, which is quite inefficient in practice. We use instead a
heuristic approach combining refinements of the isolating intervals of the real
algebraic numbers with interval arithmetic. This excludes $3^{11}-3$ of the
candidate solutions. The three remaining candidates require no further checking
since we already know that $|S_{500}|\geq3$. The overall CPU time is 71.3
seconds for 11 runs of VS plus 11 runs of CAD, followed by 16 hours for checking
candidates.\footnote{All QE-related computations have been carried out on a 2.4
  GHz Intel Core i7 with 3 GB RAM or cores on a compute server with similar
  speed and memory limitations.} Our checking procedure is a file-based
prototype starting a Reduce process for every single of the $3^{11}$ candidates;
there is considerable room for optimization.

For $k_{19}=200$ instead of $500$ all eleven univariate CAD computations yield
unique solutions which can be straightforwardly combined to one unique
solution for the corresponding $\varphi_{200}$. The overall CPU time here is
66.4 seconds for 11 runs of VS plus 11 runs of CAD. Machine float approximations of all
our solutions are given in Table~\ref{TAB:k19fix}.

\begin{table*}[t]
  \caption{The unique solution $x^{(200)}$ for $k_{19}=200$ and the three
    solutions $x^{(500)}_1$, \dots,~$x^{(500)}_3$ for $k_{19}=500$. We have actually computed
    real algebraic numbers, which are pairs of univariate polynomials and
    isolated intervals. For convenience we are giving machine float
    approximations here, which can be made arbitrarily precise.\label{TAB:k19fix}}
  \begin{align*}
x^{(200)} & =  (90.6512,2.67311,10.4996,17.8545,35.9695,32.0501,0.0954536,15.5631,2.39331,0.641001,45.4331)\\[1ex]
x^{(500)}_1 & =  (17.6392,6.97675,367.57,36.6772,5.50874,12.811,0.511775,83.4416,8.06095,0.25622,2.73253)\\
x^{(500)}_2 & =  (122.034,14.6721,234.974,14.5102,7.16952,35.064,0.42579,69.4223,7.43877,0.70128,15.2681)\\
x^{(500)}_3 & =  (323.761,9.49621,37.1013,6.72938,13.6295,43.1428,0.127807,20.8381,3.21139,0.862856,61.4581)
\end{align*}
\end{table*}

\paragraph{Parametric Analysis for $k_{19}$}
We now consider $\varphi_{k_{19}}=\varphi\sigma[100/k_{17},50/k_{18}]$ leaving
$k_{19}$ as a parameter. Again, we solve for $i\in\{1,\dots,11\}$ eleven QE
problems using VS:
\begin{displaymath}
  \varphi_{k_{19}}^{(i)}=\VS(\exists x_1\dots\exists x_{i-1}\exists
  x_{i+1}\dots\exists x_{11}\varphi_{k_{19}}).
\end{displaymath}
This time each $\varphi_{k_{19}}^{(i)}$ is a bivariate quantifier-free formula
in $k_{19}$ and the corresponding $x_i$. This time we construct a two-dimensional CAD for each $\varphi_{k_{19}}^{(i)}$. The projection order is
important: we first project $x_i$, then the CAD base phase decomposes the
$k_{19}$-axis, followed by an extension phase that decomposes the $x_i$-space
over the $k_{19}$-cells obtained in the base phase. This is feasible with one
limitation: we do not extend over zero-dimensional $k_{19}$-cells. In other
words, we accept finitely many blind spots in
parameter space, which we can explicitly read off from the CAD so that in the
end we know exactly what we are missing.

\begin{figure*}
\centering
    \includegraphics[width=\textwidth]{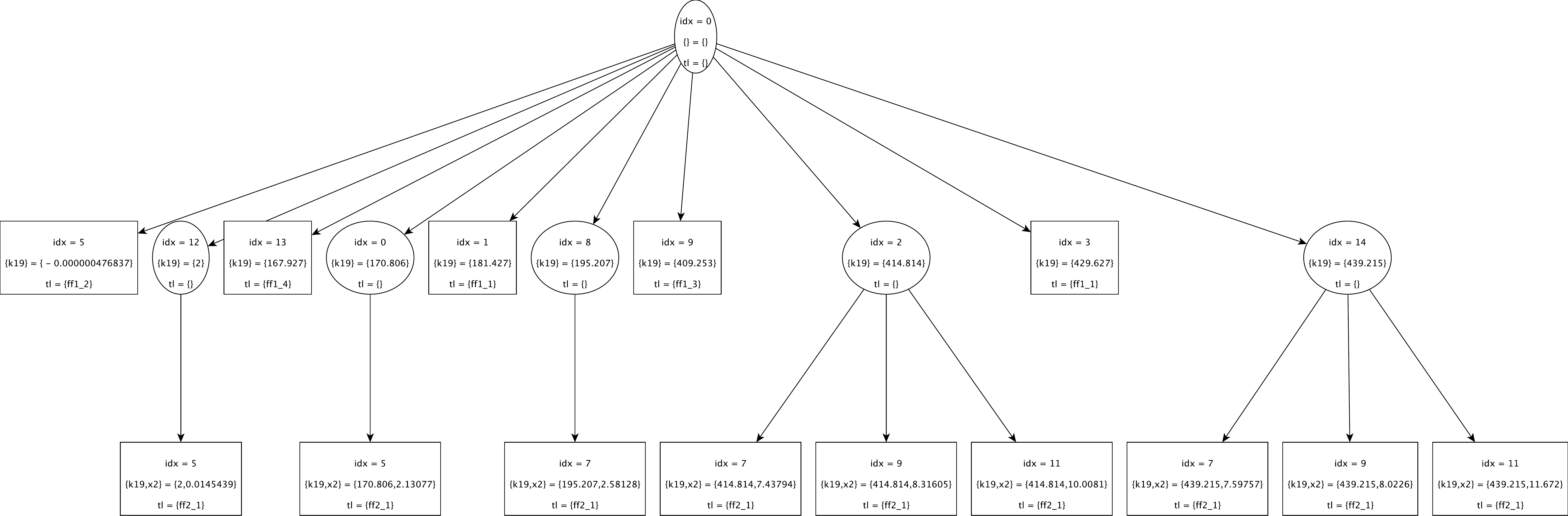}
  \caption{The pruned CAD tree for $x_2$. Ellipses and rectangles are
    full-dimensional and zero-dimensional cells, respectively. We have removed
    cells where $k_{19}$ is negative or where the input formula is
    false.\label{FIG:CAD-x2}}
\end{figure*}

Figure~\ref{FIG:CAD-x2} shows our CAD tree for $\varphi_{k_{19}}^{(2)}$. The first
layer next to the root shows the decomposition of the $k_{19}$-axis. The five
zero-dimensional (rectangular) cells are the previously mentioned blind spots,
among which the smallest one with negative value of $k_{19}$ is not relevant.
Those zero-dimensional cells also establish the limits of the full dimensional
(oval) cells in between. The cylinders over those one-dimensional
$k_{19}$-cells each contain either one or three zero-dimensional $x_2$-cells where
$\varphi_{k_{19}}^{(2)}$ holds. We have deleted from the tree all $x_2$-cells
where $\varphi_{k_{19}}^{(2)}$ does not hold. We make two observations, 
important for a qualitative analysis of our system:
\begin{enumerate}[(i)]
\item For all positive choices of $k_{19}$---extending to infinity---there is at
  least one positive solution for $x_2$.
\item There is a break point around $k_{19}=409.253$ where the system changes
  from unique solutions to exactly three solutions.
\end{enumerate}
Recall that for all floating point numbers given here as approximations we in
fact know exact real algebraic numbers. For instance, the exact break point
is the only real zero in the interval $\mathopen(409,410\mathclose)$ of an irreducible defining
polynomial
\begin{equation}
  \label{EQ:definingpol}
  \textstyle \sum_{i=0}^{10}c_ik_{19}^i\
  \text{with integer coefficients $c_i$ as in Table~\ref{TAB:coeffs}}.
\end{equation}

\begin{figure*}[p]
\centering
    \includegraphics[width=\textwidth]{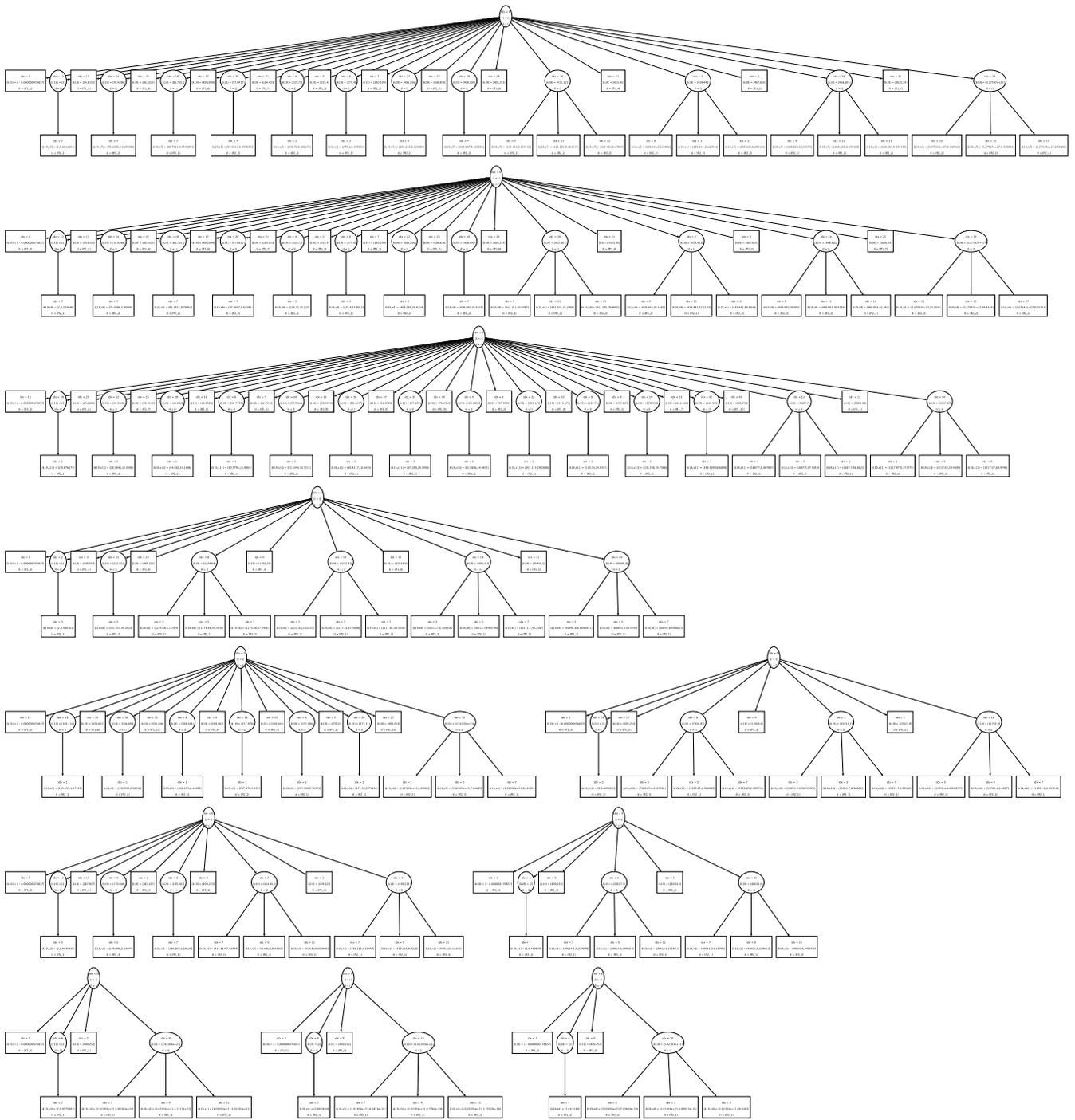}
  \caption{All CAD trees for $\psi_{k_{19}}^{(1)}$,
    \dots,~$\psi_{k_{19}}^{(11)}$. For positive $k_{19}$ there are always either
    one or
    three positive solutions for the corresponding $x_i$. The break point
    from one to three solutions is the same in all trees. In the second but last
    row on the left hand side there is the tree for $\psi_{k_{19}}^{(1)}$, which
    is shown in more detail in Figure~\ref{FIG:CAD-x2}.\newline
    \label{FIG:cad-all}}
\end{figure*}

Figure~\ref{FIG:cad-all} depicts all eleven CAD trees for $\psi_{k_{19}}^{(1)}$,
\dots,~$\psi_{k_{19}}^{(11)}$.
They
%
are quite similar to the one just discussed. Even the break point from one to
three solutions for $x_i$ is identical for all $i\in\{1,\dots,11\}$ so that we
can generalize our observations:
\begin{enumerate}[(i)]
\item For all positive choices of $k_{19}$---extending to infinity---there is at
  least one positive solution for $(x_1,\dots,x_{11})$.
\item There is a break point $\beta$ around $k_{19}=409.253$ where the system
  changes its qualitative behavior. We have exactly given $\beta$ as a real
  algebraic number in Equation~(\ref{EQ:definingpol}). For $k_{19}<\beta$ there is
  exactly one positive solution for $(x_1,\dots,x_{11})$. For $k_{19}>\beta$
  there are at least $3$ and at most $3^{11}$ positive solutions for
  $(x_1,\dots,x_{11})$.
\end{enumerate}
The overall computation time for our parametric analysis is 4.3~minutes. It is
strongly dominated by 2.8 minutes for the computation of one particular CAD tree, for
$\varphi_{k_{19}}^{(11)}$. It turns out that the
suitable projection order with $x_i$ eliminated first is computationally considerably
harder than projecting the other way round. As a preprocessing step we apply CAD-based simplification of the $\varphi_{k_{19}}^{(i)}$ with the opposite, faster, projection order. Here we use Qepcad B, which performs better than Redlog at simple solution formula construction.


\subsubsection{Triangular Decomposition methods with the Regular Chains Library}
\label{sec:LRT}

We now describe an alternative approach to the solution using regular chains methods.  Regular chains are the triangular decompositions of systems of polynomial equations (triangular in terms of the variables in each polynomial).  Highly efficient methods for working in complex space have been developed based on these; see \cite{Wang2000} 
for a survey.

Recent work by Chen et al.~\cite{CDMMXX13} proposes adaptations of these tools to the real analogue: semi-algebraic systems.  They describe two algorithms to decompose any real polynomial system into finitely many regular semi-algebraic systems.  The first does so directly while the second, Lazy Real Triangularize (LRT) produces the highest dimension solution component and unevaluated function calls, which if all evaluated would combine to give the full solution.  These algorithms are implemented in the Regular Chains Library\footnote{www.regularchains.org} in Maple which we use throughout this subsection.

We apply LRT on the quantifier-free formula (\ref{eq:varphi}) evaluated with the parameter estimates for $k_1$, \dots,~$k_{18}$ given at the start of Section \ref{secMAPK:system}, so we have one free parameter as in the previous section.
We need to choose a variable ordering: our analysis requires that $k_{19}$ be the indeterminate considered alone; the remaining variables are placed in lexicographical order (the in-built heuristics to make the choice could suggest nothing better).   The solutions must hence contain constraints in $k_{19}$, constraints in ($x_1, k_{19})$, in ($x_2, x_1, k_{19})$ and so on.  We define the \emph{main variable} of a constraint to be the highest one present in this ordering.

LRT produces one solution component and 6 unevaluated function calls in less
than 3 seconds.  In the evaluated component: for each of  $x_2$, \dots,~$x_{11}$
there is a single equation which had this as the main variable.  Further, these
are all linear in their main variable meaning they can be easily rearranged
into the solution formulae in Table \ref{tab:LRTformulae}.

\begin{table*}[t!]
  \caption{Coefficients $c_i$ and $d_j$ of polynomials occurring in 
    Equations (\ref{EQ:definingpol}) and (\ref{eq:x1pol}), respectively.\label{TAB:coeffs}}
  {\small
\begin{align*}
  c_{10} &= \relax
             351590934502740290936895033267017158736060313940693076650155371250411\\
  c_9    &= -213699072852157674283997527746395583273033983170426080574800781989093156\\
  c_8    &= \relax
             25374851641220554774259605635053469432582109883965015804077119110958034090\\
  c_7 &= \relax12972493018300022707027639267804259251235991618029852880330004508564391594000\\
  c_6 &= -8468945963692802414226427249726123493448372439778349029355636316929687020660000\\
  c_5 &= \relax2231098270337406450670301663172664333421440833875848621423683265663846533079600000\\
  c_4 &= -376265008904112258290319173193792052014899485528994925965885895511831873444245100000\\
  c_3 &= \relax39262101548790869407057994985320156500968958361396178908180026842806643766783104000000\\
  c_2 &= -2492623990743029234974354081270296106309603462451517057779877596842448287799337600000000\\
  c_1 &= \relax70978850735887473459176997186175978425873267246760023212940616924643171868478080000000000\\
  c_0 &= -1062871192838985876948077114923898204990434138901495394834749613184670362810368000000000000\\[2ex]
  d_6 &= 16838105723097694257603469\\
  d_5 &= -24078605201553273505077988k_{19} + 7723967969644977896148686580\\
  d_4 &= 8176202638735769127032169k_{19}^2  - 7723411665463544477701499460k_{19} + 1232154357941338876156606812900\\
  d_3 &= 1465408757440589841803452380k_{19}^2  -
        798169557586805582842481309800k_{19}  +
        83152655240002767729550477640000\\
  d_2 &= 85462524901276846107251669400k_{19}^2  - 35266411401427656834572095140000k_{19}  + 2556805354853318332197489636000000\\
  d_1 &= 1631685649719702672282505500000k_{19}^2  - 721989571100461862477342320000000k_{19}  + 28843755938318780823218400000000000\\
  d_0 &= -7013104139459910876520500000000000k_{19}.
\end{align*}}
\end{table*}

\begin{table*}[t!]
  \caption{Triangular solution formulae valid for all positive $k_{19}$
    excluding three isolated points
}
\label{tab:LRTformulae}
\begin{align*}
x_{11} &= -\frac{1}{60}x_{2}^2 
+ \frac{1}{600}(10k_{19} - 10x_{1} - 37x_{3} + 10x_{4} - 2100)x_{2}
-\frac{9}{200}x_{3}^2 + \frac{1}{600}(-27x_{1} + 27x_{4} + 27k_{19} - 4650)x_{3}
\\ & \phantom{={}} {}- x_{1} + x_{4} + k_{19} - 50
\\[\lrtskip]
x_{10} &= \frac{1}{150}x_{2}(x_{2}+x_{3}-x_{4}-k_{19}+x_{1}+150)
\\[\lrtskip]
x_{9} &= \frac{1}{18200}(69x_{3}+182x_{2})(x_{2}+x_{3}-x_{4}-k_{19}+x_{1}+150)
\\[\lrtskip]
x_{8} &= \frac{15}{364}(x_{2}+x_{3}-x_{4}-k_{19}+x_{1}+150)x_{3}
\\[\lrtskip]
x_{7} &= 50 - \frac{2}{101}x_{4}x_{1} - x_{4}
\\[\lrtskip]
x_{6} &= \frac{2}{101}x_{4}x_{1}
\\[\lrtskip]
x_{5} &= x_{2} + x_{3} - x_{4} - k_{19} + x_{1} + 150
\\[\lrtskip]
x_{4} &= \frac{2525000}{101x_{2}+1000x_{1}+50500}
\\[\lrtskip]
x_{3} &= \frac{-101x_{2}^3 - (-101k_{19} + 1101x_{1} + 65650)x_{2}^2 - (1000x_{1}^2 + (-1000k_{19}+200500)x_{1}  - 50500k_{19}+5050000)x_{2}  + 150000x_{1})}{101x_{2}^2 + (1000x_{1}+50500)x_{2}}\\
  x_{2} &= \frac{n}{d} \quad \mbox{where}\quad\vtop{\halign{
          $#{}$ & ${}#$\hfill\cr
                  n = &  30625833064790009548991419920x_{1}^5 + (-43795148662369306906962603840k_{19}\cr
      & + 37749979225487731805273686504663200)x_{1}^4 + (14871210647782462053693235920k_{19}^2 \cr
      & - 16963336293692750919154910690672400k_{19} + 6815925407229297763234036009365120000)x_{1}^3 \cr
      & + (1538325448222983229930530049200k_{19}^2 - 862702164104208291031357996000020000k_{19}\cr
      & + 279241219028720368578809336249748000000)x_{1}^2 + (29370341694954648101085099000000k_{19}^2 \cr
      & - 12995812279808313524592161760000000k_{19} + 3705960282117523242886769213700000000000)x_{1}\cr
      & - 126235874510278395777369000000000000k_{19}\cr
        \noalign{\vskip\lrtskip}
        d = & 232763663752113237974029404420089x_{1}^5 + ( - 332853615301041845577671639990228k_{19} \cr
      & + 88646303215205075376308147029677220)x_{1}^4 + (113024761399450186949390623074789k_{19}^2\cr
      & - 80843908028331498139954527761762740k_{19} + 11682465068391769796632986929072776500)x_{1}^3\cr
      & + (11455232309649034305597048791479020k_{19}^2 - 5547251026060433566640620528023877000k_{19} \cr
      & + 619147207587597001268026254404647600000)x_{1}^2 + (290245997063001550130198026458525000k_{19}^2 \cr
      & - 141348286758352762323489548674398500000k_{19}\cr
      & + 14547288529581382252587071541494600000000)x_{1}\cr
      & - 1247498501818579946626756931775000000000(k_{19}-100)\cr
}}\\[\lrtskip]
x_1 & \phantom{{}={}} \text{has at most 6 solutions for a given value of $k_{19}$, according to Equation (\ref{eq:x1pol}).}
\end{align*}
\end{table*}

The constraints on $(x_1, k_{19})$ are that $x_1>0$ and that a polynomial equation of degree $6$ be satisfied:
\begin{equation}
\label{eq:x1pol}
f(x_{1}, k_{19}) = \textstyle \sum_{i=0}^{6} d_ix_{1}^i = 0
\end{equation}
where the coefficients $d_i$ are univariate polynomials in $k_{19}$ of maximum degree $2$ as given in Table
\ref{TAB:coeffs}.


Finally, the constraints on $k_{19}$ are that it be positive; it not be a root of the polynomial in Equation (\ref{EQ:definingpol}); nor two other polynomials as described in Table \ref{tab:k19constraint}.
\begin{table*}[t!]
\caption{Constraints on $k_{19}$ for solution formulae in Table \ref{tab:LRTformulae} and Equation (\ref{eq:x1pol}) to be valid}
\label{tab:k19constraint}
\begin{align*}
  k_{19}&>0\\[\lrtskip]
  \mbox{polynomial in (\ref{EQ:definingpol})} & \neq 0\\[\lrtskip]
  23197989433419579994929k_{19}^2 
  - 89407400615452409453098800k_{19}
  - 4822419303419166525491149190000 &\neq 0\\[\lrtskip]
  505465566622475867655547880786544637953790406059982726185509k_{19}^4\\
  {}-  12725780456964391893178560515183873684222178969868366920505134120k_{19}^3\\
  {}+ 1175510330915205241831243213231417517003037315562884193657451445400k_{19}^2\\
  {}- 281867359883676159811192082978541193600292804324596911878337972560000k_{19}\\
  {}- 42434363570215587465668423701563932185051066892741207931879307200000000 & \neq 0
\end{align*}
\end{table*}


Thus this solution component is valid for all positive values of $k_{19}$ excluding three points.  As before, we could give these as exact algebraic numbers but for brevity give float approximations: $409.253$, $16473.337$, and $25084.536$.

Three of the six unevaluated function calls define the solutions at these points, however evaluating these solutions is not possible in reasonable time.  The other three define empty solution sets (evaluating to discover this is instantaneous).  So, as with our previous approach, we proceed accepting a small number of blind spots.  

The output of LRT has quickly given us the structure of the solution space valid at all but three isolated values of $k_{19}$.  However, it does not identify where the number of real solutions change: although the break point identified earlier has been rediscovered there is no information from which we can infer its significance; and there is no significance in our application of the other two isolated points.  

To finish the analysis we need to decompose $(x_1, k_{19})$-space according to the real roots of $f(x_1, k_{19})$; and also $x_1$ since the constraint $x_1>0$ was specified separately in the output (the case for this variable only).  CAD is ideally suited for this task. Using the Regular Chains algorithm \cite{CMXY09} in Maple a CAD for $f(x_{1}, k_{19})$ 
divides the plane into 135 cells in a few seconds.  This CAD decomposes the $k_{19}$ axis into 11 cells, i.e. identifying five points which approximate to: $-379.993$, $-87.776$, $0$, $409.253$, and $25084.536$. 

On the cell for $k_{19} \in \mathopen]0, 409.253\mathclose[$, the cylinder above in the $(x_1, k_{19})$ plane is divided into 11 cells: three of which cover $x_1>0$ (two 2d sectors and a 1d section).  This indicates that $f(x_1, k_{19})$ has a single positive real solution for such $k_{19}$.  On the two cells for $k_{19} \in \mathopen]409.253, 25084.536\mathclose[$ and $k_{19} \in \mathopen]25084.536, \infty\mathclose[$ the cylinders above are divided into 15 cells; seven of which cover $x_1>0$.  This indicates that $f(x_1, k_{19})$ has three positive real solutions for such $k_{19}$.

At the end of this analysis we have rediscovered the break point where the system moves from a single positive real solution to three.  We also have explicit solutions valid for all except three isolated $k_{19}$ values.  To obtain a solution select the $k_{19}$ value of interest then identify the real roots of $f(x_1, k_{19})$ (we know in advance how many depending on the $k_{19}$ value chosen); then for each $x_1$ solution substitute recursively into the equations of Table \ref{tab:LRTformulae};  starting from the bottom and including the new variable solution discovered from each substitution into the next.  The solutions in Table \ref{TAB:k19fix} may be easily rediscovered this way.

\paragraph{Repeating the Process for Different Choices of the Lone Free Parameter/Fixed Parameter Values}

We may repeat the approach described above for different choices of free parameter and different choices of fixed parameter values.  For example:
\begin{itemize}
\item With $k_{17}$ set to $95$ instead of $100$ we find that the break point between 1 and 3 real positive solutions moves to $k_{19} = 369.917$.  With $k_{17}$ set to $105$ it moves to $k_{19} = 450.077$.
\item Allowing $k_{17}$ to be free and fixing $k_{19} = 200$ we find that there is only ever one positive real solution.
\item Allowing $k_{17}$ to be free and fixing $k_{19} = 500$ we find the number of positive real solutions moving from 1 to 3 to 1 breaking at $k_{17} = 85.988$ and $k_{17} = 110.869$.  
\item Similarly, allowing $k_{18}$ to be free and fixing $k_{19} = 200$ we find there is only ever one positive real solution; but fixing $k_{19}=500$ instead we find 3 real solutions between $k_{18} = 51.382$ and $58.329$ and 1 otherwise.  
\end{itemize}
The results above hint that there is a shape approximating a narrow paraboloid in $(k_{17}, k_{18}, k_{19})$-space within which bistability may occur; with bistability available for any $k_{17}$ and $k_{18}$ value but bounded from below in the $k_{19}$ coordinate.
We note that these additional experiments all produce results which, as with the one described in detail, are invalid at a handful of isolated values of the free parameter.  

\subsection{Stability of the Fixed Points}

We use the three linear conservation constraint equations (\ref{EQ:claws})
to eliminate $x_1$, $x_7$, and $x_{11}$ from system (\ref{EQ:thesystem}) and symbolically compute the Jacobian $\tilde{J}$ of the obtained reduced system. We then numerically compute the eigenvalues of $\tilde{J}$ for the instances arising from the substitution of the different positive fixed points for the variables and the corresponding parameter values. 

We have used the float approximations for the unique solution $x^{(200)}$ with $k_{19}=200$ and the three solutions $x^{(500)}_1$, \dots,~$x^{(500)}_3$ for $k_{19}=500$ in Table \ref{TAB:k19fix}. 
For the single positive fixed point $x^{(200)}$ the Jacobian $\tilde{J}(x^{(200)})$ has eigenvalues with negative real part only and hence can be shown to be stable. For $k_{19}=500$ one of the three positive fixed points $x^{(500)}_2$ can be shown to be unstable, as $\tilde{J}(x_2^{(500)})$ has one eigenvalue with positive real part; the other seven had negative real parts.  In contrast $x_1^{(500)}$ and $x_3^{(500)}$ can be shown to be stable. Hence for $k_{19}=500$ the system is indeed bistable.

A verification of the stability of the fixed points using exact real algebraic numbers by the well-known Routh--Hurwitz criterion
is possible algorithmically \cite{HongLiskaSteinberg97}, but 
seems to be out of range of current methods for this example.
Notice that also in other studies on multistationarity of signalling pathways \cite{Conradi2008,Gross2016} 
the question of stability has not been addressed.

\subsection{Numerical Homotopy Methods} 

Finally, we compare our symbolic results with numerical ones obtained using the homotopy solver Bertini \cite{BHSW06}.  Bertini computes complex roots of polynomial systems using methods from numerical algebraic geometry \cite{Sommese2005}. 

For the parameter values as above and $k_{19}=500$ we obtain six
solutions, three of which are positive real solutions. For $k_{19}=200$ we obtain a single positive solution. In both cases the relevant solutions coincide with the ones obtained with our symbolic analyses up to the used numeric precision.

However, for larger values of $k_{19}$ Bertini produces incorrect results due to numerical instability. For instance, we falsely obtain exactly one positive real solution for $k_{19}=6000$ and no positive real solution for $k_{19}=10000$.

\begin{figure*}[t]
\centering
    \includegraphics[width=0.99\textwidth]{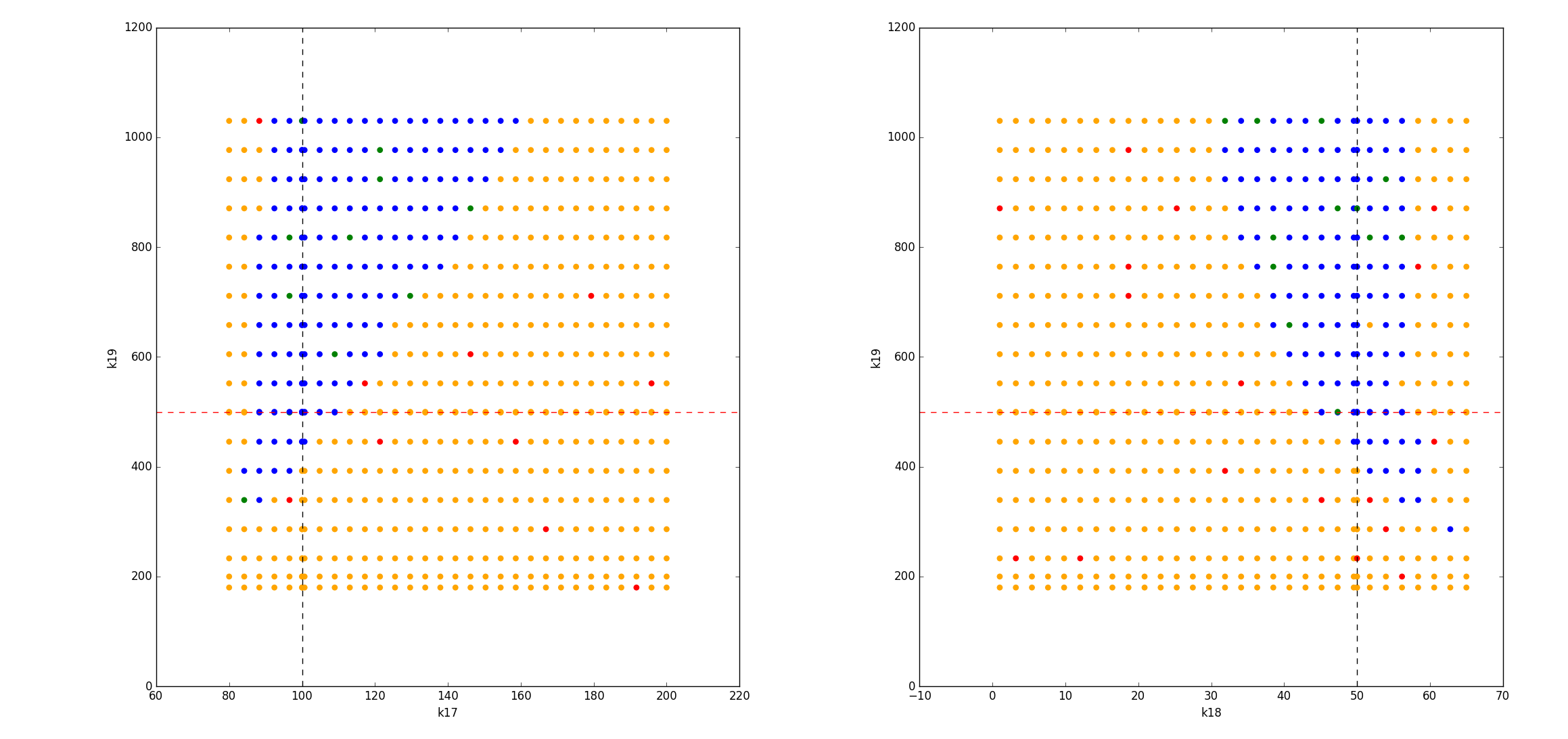}
  \caption{Grid sampling two-parameter regions using Bertini. We combine
    $k_{19}$ with $k_{17}$ (left) and with $k_{18}$ (right). Colors indicate the
    computed numbers of positive real fixed points: blue 3, green 2, yellow 1,
    red 0. The dotted lines indicate values of the parameters as in Equation~(\ref{EQ:clestimates}).\label{FIG:uniform_k17_k18}}
\end{figure*}

Figure~\ref{FIG:uniform_k17_k18} shows a Bertini-based grid sampling of parameter regions, varying $k_{19}$ between 200 and 1000 and fixing one of $k_{17}$ and $k_{18}$ while varying the other among the default values (\ref{EQ:clestimates}).  While this suffers from the discrete nature of sampling and potentially unreliable results as discussed, it is nevertheless useful for the generation of hypothesis about the nature of the parameter regions. 
Figure \ref{FIG:uniform_k17_k18} seems to identify a region of bistability (in blue) within the parameter space, as hypothesised at the end of Section \ref{sec:LRT}.  The results of Bertini indicate holes in this region (the green dots within the blue).  However, computation at these particular points reveals these to be the result of numerical errors:  where an insufficiently high precision causes what is actually a positive real solution to appear to have a negative component.  It seems there is scope for fruitful interplay between symbolic and numeric methods here; with numerics postulating hypotheses for the symbolic methods to check and refine.

\section{Conclusions and Future Work}

We have shown that the determination of multistationarity of an 11-dimensional MAPK network can be achieved by combinations of currently available symbolic computation methods for mixed equality/inequality systems if, for all but potentially one parameter, numeric values are known. 
The aspiration of a semi-algebraic description of the ranges for all parameters in the conservation laws (\ref{EQ:claws}) yielding multistationarity will now be pursued, with the present results demonstrating that this aspiration may be within reach.


As there are many very relevant systems having dimensions between 10 and 20
it seems to be worth the effort to enhance and improve the present algorithmic methods, and in particular their combination, to solve such important application problems
for symbolic computation. 


\section*{Acknowledgments}
For our QE-related computations we used two great free software tools:
GNU Parallel for distributing computations on several processors, and
yEd for visualization of CAD trees. D.~Grigoriev is grateful to the
grant RSF 16-11-10075 and to MCCME for wonderful working conditions
and inspiring atmosphere. M.~Ko\v sta has been supported by the
DFG/ANR Project STU 483/2-1 SMArT.
H.~Errami, A.~Weber, and O.~Radulescu thank the
French-German Procope-DAAD program for partial support of this
research. J.~Davenport, M.~England and T.~Sturm are grateful to EU CSA project
712689 SC\textsuperscript{2}.

\bibliographystyle{abbrv}
\bibliography{issac2017}
\end{document}